\documentclass[11pt]{article}
\usepackage[latin1]{inputenc}
\usepackage[T1]{fontenc}
\usepackage[ngerman]{babel}
\usepackage{a4}

\usepackage{amsmath}
\usepackage{amssymb}
\usepackage{amsthm}
\usepackage{times}
\usepackage{graphicx}
\usepackage{wrapfig}
\usepackage{color}
\usepackage{rotating}

\begin{document}

\title{Gravitation, "Aquivalenzprinzip und Quantenmechanik}
\author{Domenico Giulini\\
Zentrum f"ur angewandte Raumfahrttechnologie und Mikrogravitation, Bremen\\
Institut f"ur Theoretische Physik der Leibniz Universit"at Hannover}
\date{}
\maketitle

\begin{abstract}
\noindent
Die Gravitation ist gem"a"s der Allgemeinen Relativit"atstheorie 
ein Merkmal der Geometrie von Raum und Zeit und damit keine Kraft 
im Newton'schen Sinne mehr. Dies ist eine Folge des Einstein'schen
"Aquivalenzprinzips, das bisher in allen experimentellen 
Pr"azisionstests best"atigt wurde. Die Suche nach m"oglichen
Verletzungen dieses Prinzips h"alt aber unvermindert an, da 
diese gegebenenfalls wertvolle Hinweise auf weiterf"uhrende 
Theorien geben k"onnten. Insbesondere wird erwartet, dass das 
Studium der Einwirkungen von Gravitationsfeldern auf genuin nur 
quantenmechanisch zu beschreibende Systeme, wie Atome und Molek"ule, 
hier neue M"oglichkeiten er"offnet. Das wirft wiederum einige 
fundamentale Probleme auf, da Allgemeine Relativit"atstheorie 
und Quantenmechanik auf teilweise inkompatiblen Begriff-Systemen 
basieren.  In diesem Artikel, der in gek"urzter Form in der 
Zeitschrift \emph{Spektrum der Wissenschaft} erscheinen wird, 
werden die damit verbundenen Aspekte beleuchtet und anhand 
eines in den letzten Jahren viel diskutierten Beispiels 
konkretisiert. 
\end{abstract}

% Gravitation, according to General Relativity, is an attribute of
% space-time's geometry and hence not a force in the Newtonian sense.
% This is a consequence of Einstein's equivalence principle, which 
% so far passed all experimental tests with high precision. However, the  
% search for possible violations continues, for they, as the case
% may be, are expected to point towards more fundamental theoretical
% extensions of General Relativity. In particular, it is expected that 
% useful insights are gained by studying the interaction between 
% gravitational fields and genuine quantum-mechanical systems, like 
% atoms or molecules. But this raises some fundamental issues, for 
% General Relativity and Quantum Theory rest on partially incompatible 
% sets of concepts. This article tries to explain in an elementary fashion 
% how these issues recently entered modern research in atom interferometry. 
% A shorter and editorially adapted  version of this paper will appear 
% in ``Spektrum der Wissenschaft'' -- the German edition of 
% ``Scientific American''. 

\section{Einleitung}
Erkl"artes Ziel wissenschaftlicher Naturerkl"arung sind Theorien, 
aus denen das tats"achlich beobachtete Naturgeschehen deduktiv 
aus m"oglichst wenigen und einfachen Prinzipien abgeleitet werden 
kann. Obwohl dies wegen des letztlich immer mit Unsicherheiten 
behafteten Wirklichkeitsbezugs nicht zu einem makellosen 
System von Axiomen im Sinne der Mathematik f"uhren kann, ist es 
dennoch m"oglich, fundamentalere Strukturen von weniger
grundlegenden zu unterscheiden. Dabei werden Theorien oder 
Theorieteile im Sinne des eben genannten Ziels als fundamental 
angesehen, wenn sie hinreichend gro"se Klassen von 
Ph"anomenen unter Prinzipien von hinreichend allgemeiner 
Anwendbarkeit ordnen. Dabei soll die Anwendbarkeit m"oglichst 
eindeutig und in ihren Konsequenzen nat"urlich nicht 
widerspr"uchlich sein, wof"ur eine widerspruchsfreie 
mathematische Ordnung die notwendige (aber nicht hinreichende) 
Bedingungen darstellt. 

Beispiele f"ur Theorien, die als fundamental angesehen werden 
k"onnen, sind die Spezielle und Allgemeine Relativit"atstheorie, 
sowie die Quantentheorie. Sinnbildlich f"ur deren grundlegenden 
Charakter steht die Tatsache, dass jede dieser Theorien 
wesentlich mit der Existenz einer Naturkonstanten zusammenh"angt: 
F"ur die Spezielle Relativit"atstheorie ist dies die 
Lichtgeschwindigkeit im Vakuum $c$, f"ur die Allgemeine 
Relativit"atstheorie die Gravitationskonstante $G$ und f"ur 
die Quantentheorie das Plancksche Wirkungsquantum $h$. 
Diese Gr"o"sen sind dimensionsbehaftet und lassen sich 
durch Kombinationen der Einheiten von L"ange, Zeit 
und Masse ausdr"ucken. Umgekehrt k"onnen durch algebraische 
Kombinationen von $c,G$ und $h$ Einheiten f"ur L"ange, 
Zeitdauer und Masse rekonstruiert werden, die man die 
\emph{Planck-Einheiten} nennt und die in Anhang~1 
n"aher erl"autert sind. Sie geben einen ersten Hinweis darauf, 
unter welchen physikalischen Bedingungen wohl alle drei Theorien 
gleicherma"sen wichtig werden. 

Ungl"ucklicherweise sind aber die Planck'sche L"ange und Zeitdauer 
so aberwitzig klein, dass es ausgeschlossen erscheinen mag diese 
experimentell auch nur ann"ahernd aufzul"osen. So liegen zwischen 
der Planck L"ange von etwa $10^{-34}$ Metern und der Dimension 
eines Korns feinem Sand (0,1 Millimeter) genauso viele Gr"o"sen\-ordnungen 
(Zehnerpotenzen) wie zwischen der Gr"o"se dieses Korns 
und dem Durchmesser des gesamten sichtbaren Universums 
(etwa $10^{26}$ Meter). Allerdings darf man hier nicht zu 
schnell in einen naiven Operationalismus verfallen. Als warnendes 
Beispiel sei angef"uhrt, dass heutige Gravitationswellen-Detektoren 
auf laser-interferometrischer Basis relative L"angen"anderungen 
von $\Delta L/L\approx 10^{-21}$ messen k"onnen, was umgerechnet 
auf die Basisl"ange von etwa einem Kilometer einer absoluten 
L"angen"anderung von $10^{-18}$ Metern entspricht, also einem 
Tausendstel des Protonenradius! Wer bei L"angen nur an materielle 
Ma"sst"abe denkt w"urde einer solchen Aussage wohl schnell 
ihre physikalische Sinnhaftigkeit absprechen; und doch hat 
sie in der Laseroptik einen wohl definierte Bedeutung! 
Entsprechend vorsichtig sollte man auch mit Behauptungen 
"uber eine angebliche Unerreichbarkeit der Planck-Skalen sein.

Aber auch ohne in das ambitionierte Gesch"aft einer Quantengravitation 
einzusteigen kann man sich nach den physikalischen und begrifflichen 
Grundlagen des Wechselspiels zwischen quantentheoretisch beschriebener 
Materie und gew"ohnlich (klassisch) beschriebener Gravitation fragen. 
Wenn sich das Verhalten eines makroskopischen K"orpers aus dem Verhalten 
seiner atomaren Konstituenten ergibt, muss sich auch die gravitative 
Wechselwirkung auf atomarer Ebene beschreiben lassen. Aber f"allt ein 
Atom immer so wie ein Stein? Und was bedeutet diese Frage, wenn sich das 
Atom in einem r"aumlich nicht lokalisierten Superpositionszustand befindet, 
der nach der Quantenmechanik ja nicht nur m"oglich sondern im Labor 
auch gezielt herstellbar ist? Gibt es "`erste Prinzipien"', aus denen 
wir ableiten k"onnen, wie die Gravitation auf quantenmechanisch 
beschriebene Materie einwirkt?  

\section{Die drei S"aulen des Einstein'schen "Aquivalenzprinzips}
Das Einstein'sche "Aquivalenzprinzip scheint nun genau das zu sein, 
was wir brauchen. Es ist ein so genanntes heuristisches Prinzip, 
dient also mit mehr oder weniger scharf gegebenen Anweisungen 
der Auffindung neuer Gesetzm"a"sigkeiten; hier der Ankoppelung
(Wechselwirkung) von Materie an Gravitation. Ph"anomenologischer Ursprung 
dieses Prinzips ist die "uberraschende Gleichheit von schwerer (d.h. 
f"ur die Gravitationswirkung verantwortlicher) und tr"ager (d.h. f"ur 
das Tr"agheitsverhalten verantwortlicher) Masse eines jeden K"orpers, 
die vor Einstein bereits von Heinrich Hertz in seiner halb-popul"aren 
Vorlesung \emph{Die Constitution der Materie} des Jahres 1884 pr"agnant 
zusammengefasst wurde. Das Manuskript dieser Vorlesung, das erst 
nach dem Tode von Hertz' j"ungeren Tochter Mathilde in deren 
Nachlass durch den Wissenschaftsjournalisten und Hertz-Biographen 
Albrecht F"olsing wiederentdeckt wurde und mittlerweile in Buchform 
erh"altlich ist~\cite{Hertz:ConstitutionDerMaterie}, enth"alt u.A. 
folgenden Passus:
 
\bigskip 
\begin{wrapfigure}[22]{r}{0.5\linewidth}
\centering\includegraphics[scale=0.45]{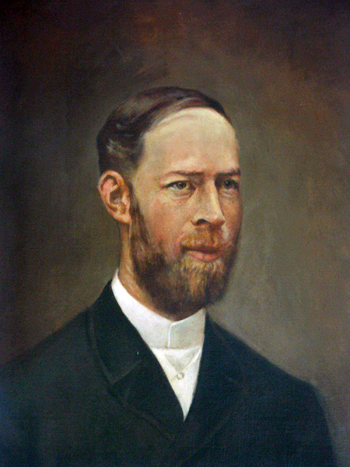}
\caption{\small Heinrich Hertz (1857-1894) thematisierte schon 1884 das
Problem der Gleichheit von schwerer und tr"ager Masse.}
\end{wrapfigure}
\noindent \textit{
"`Sehr wunderbar ist der Zusammenhang, welcher zwischen der 
Gravitation der Materie und ihrer Tr"agheit besteht. 
Wir sehen n"amlich, da"s irgend zwei Mengen von Materie, 
welche gleiche Tr"agheit besitzen, auch gleiche Gravitationswirkung 
aus"uben, einerlei, welches der Stoff ist, aus dem sie bestehen. [...]\\
Und doch haben wir in Wahrheit zwei Eigenschaften, zwei 
Haupteigenschaften der Materie vor uns, die v"ollig unabh"angig 
voneinander gedacht werden k"onnen und die sich durch die Erfahrung
und nur durch diese als v"ollig gleich erweisen. Diese 
"Ubereinstimmung ist also vielmehr als ein wunderbares R"athsel 
zu bezeichnen, sie bedarf einer Erkl"arung; wir d"urfen vermuthen, 
da"s auch eine einfache und verst"andliche Erkl"arung m"oglich ist, 
und da"s uns diese Erkl"arung einen weitgehenden Einblick in die 
Constitution der Materie gestatten wird."'}

Einstein gab diese Erkl"arung 30 Jahre sp"ater in Form seiner Allgemeinen 
Relativit"atstheorie, in der die "`Wesensgleichheit"' (Einstein) 
von Gravitation und Tr"agheit ihren Ausdruck in der 
vereinheitlichten geometrischen Beschreibung von Raum und Zeit 
findet. Die Raumzeit tr"agt gewisse geometrische Attribute, die 
nicht wie in der Newton'schen Physik unabh"angig vom materiellen
Geschehen ein f"ur alle mal fest vorgegeben sind, sondern 
dynamisch vom materiellen Geschehen beeinflusst werden und 
umgekehrt auch dynamisch auf die Materie zur"uckwirken. 
Dieses Wechselseitigkeit der Beeinflussung ist vom 
physikalischen Gesichtspunkt viel befriedigender als die 
einseitige Wirkung die Newtons absoluter Raum aus"ubt, der 
gegen jede R"uckwirkung der Materie immun sein soll. Sie hat 
aber ihren Preis in der erheblich komplizierteren Bauart 
der wechselseitig gekoppelten Gleichungen, die die Dynamik 
von Gravitation und Materie gemeinsam beschreiben und auch nur 
gemeinsam gel"ost werden k"onnen.

Diese geometrische Auffassung der Gravitation ist aber nur 
dann m"oglich, wenn alle Materieformen ausnahmslos die gleiche 
Geometrie ``sehen'', d.h. auf Gravitation in einer \emph{universellen} 
Weise reagieren. G"alte dies nicht, m"usste man der Raum-Zeit 
mehrere Geometrien gleichzeitig zuschreiben, die wechselweise 
ja nach der zur Vermessung benutzten Materie zur Anwendung 
kommen m"ussten. Dies w"are jedoch ein reichlich willk"urliches 
Verfahren ohne besonderen Erkl"arungswert. 

Wie genau die Universalit"at der Gravitation beschaffen 
sein muss, um durch eine eindeutige Geometrie der Raum-Zeit 
erkl"art werden zu k"onnen, ist gerade im Einstein'schen 
"Aquivalenzprinzip formuliert. Allerdings gab Einstein 
diesem Prinzip mehrere Wendungen, die streng genommen nicht 
"aquivalent zueinander sind. Deshalb hat sich heute eine 
vereinheitlichte Sprechweise herausgebildet, die genau auf 
das erkl"arte Ziel einer Geometrisierung zugeschnitten ist. 
In dieser Formulierung zerf"allt das Prinzip in folgende 
drei Teile.
\begin{itemize}
\item[1.]
\textbf{Die Universalit"at des freien Falls (UFF):}
Eine Testmasse  bewegt sich im Gravitationsfeld
auf einer Bahn, die nur vom Anfangsort und 
Anfangszeitpunkt, sowie der Anfangsgeschwindigkeit 
abh"angt. Weitere Abh"angigkeiten, wie etwa von 
der chemischen Zusammensetzung, sollen nicht existieren.
Insbesondere sollen also zwei Testmassen $A$ und $B$, 
die am gleichen Ort und zur gleichen Zeit relativ 
zueinander ruhend in einem Gravitationsfeld fallen 
gelassen werden die gleiche Beschleunigung relativ 
zu einem im Gravitationsfeld ruhenden Beobachter (Labor) 
erfahren. M"ogliche Verletzungen dieses Prinzips werden 
deshalb meist durch eine Zahl $\eta$ parametrisiert, die 
den Betrag der Differenz der Beschleunigungen (hier immer
positiv gerechnet) geteilt durch deren Mittelwert angibt:
\begin{equation}
\label{eq:EtvoesParameter}
\eta=2\cdot\frac{\vert a(A)-a(B)\vert}{a(A)+a(B)}\,. 
\end{equation}
Diese Zahl ist nach dem ungarischen Physiker Baron 
Lor{\'a}nd (Roland) von E"otv"os  (1848-1919) benannt, der 
in den Jahren 1906-09 zusammen mit Kollegen 
Pr"azisionsexperimente zur Gleichheit von schwerer 
und tr"ager Masse verschiedener Materialien (darunter 
Platin und Mangan) anstellte und nach heutiger (etwas 
vorsichtigerer) Ansicht m"ogliche Abweichungen durch 
$\eta<10^{-8}$ begrenzen konnte. Zum gegenw"artigen 
Zeitpunkt sind obere Schranken im Bereich $10^{-13}$ 
an $\eta$ f"ur einige Materialkombinationen 
bekannt. So konnte die "`E"ot-Wash"'-Gruppe an der 
Universit"at von Washington in Seattle (USA), 
%
%% Homepage www.npl.washington.edu/eotwash/
%
die seit vielen Jahren f"uhrend Pr"azisionsexperimente mit 
Torsionswaagen durchf"uhrt, den E"otv"os Parameter 
f"ur die Materialkombination Beryllium-Titan im 
Gravitationsfeld der Erde einschr"anken durch 
(Fehlerangaben entsprechen stets einer Standardabweichungen,
d.h. 68\% Konfidenz) 
\begin{equation}
\label{eq:WEP-ObereSchranke-BeTi}
\eta(\text{Be},\text{Ti})=(0.3\pm 1.8)\times 10^{-13}\,.
\end{equation}
Kennt man die Gegebenheiten des Experiments, so kann man 
diese Zahl durch die Absolutwerte der Beschleunigungen 
ausdr"ucken. Es ergibt sich, dass eventuell vorhandene 
Unterschiede in den Beschleunigungen der betrachteten Proben 
von Beryllium und Titan im Gravitationsfeld der Erde  
kleiner sind als $4\times 10^{-15}m/s^2$. 
F"ur die Zukunft sind weitere Experimente in Planung,
um m"oglichen UFF-Ver\-letzungen auf die Spur zu kommen. 
Am weitesten fortgeschritten in der Planung ist die f"ur 
2016 vorgesehene MICROSCOPE--Mission (Micro-Satellite 
\`a tra\^{\i}n\'ee Compens\'ee pour l'Observatrion du 
Principe d'Equivalence), an der neben den franz"osischen 
auch deutsche Forschungsinstitutionen beteiligt sind.%
\footnote{
CNES (Centre National d'\'Etudes Spatiales), 
OCA (Observatoire de la Cote d'Azur),
ONERA (Office National d'\'Etudes et de Recherches A\'erospatiales), 
DLR (Deutsches Zentrum f"ur Luft und Raumfahrt),
PTB (Physikalisch Technischen Bundesanstalt), 
ZARM (Zentrum für Angewandte Raumfahrttechnologie und Mikrogravitation).} 
Hier werden Relativbeschleunigungen zweier frei fallender 
Testmassen aus einer Platin-Rhodium Legierung bzw. aus 
Titan untersucht, die sich innerhalb eines Mikrosatelliten 
in einer polaren, sonnensynchronen Erdumlaufbahn in etwa 
700 Kilometern H"ohe befinden. Die prognostizierte Genauigkeit 
liegt bei $\eta\approx 10^{-15}$, stellt also eine Verbesserung 
der bisherigen Genauigkeit um zwei Gr"o"senordnungen dar.
Zum Schluss sei erw"ahnt, dass der E"otv"os-Faktor f"ur das 
Paar Erde-Mond bez"uglich ihres Falls im Gravitationsfeld 
der Sonne eine ganz "ahnlich niedrige experimentelle Schranke
besitzt,  n"amlich
\begin{equation}
\label{eq:WEP-ObereSchranke_ErdeMond}
\eta(\text{Erde},\text{Mond})=(-1.0\pm 1.4)\times 10^{-13}\,.
\end{equation}
Im Unterschied zu K"orpern im Labor ist bei astronomischen Objekten 
der Anteil der gravitativen Bindungsenergie an der Gesamtenergie 
nicht mehr zu vernachl"assigen. F"ur den Mond liegt dieser etwa 
bei $10^{-11}$, bei der Erde zwanzig mal h"oher. W"urde die 
gravitative Bindungsenergie zur tr"agen Masse wesentlich anders 
beitragen als zur schweren, h"atte sich dies im E"otv"os-Faktor
des Erde-Mond-Systems zeigen m"ussen. 
\item[2.]
\textbf{Lokale Lorentz-Invarianz (LLI):} Lokale physikalische 
Experimente weisen keine bevorzugte Raumrichtung und 
keinen bevorzugter Bewegungszustand aus. Alle frei 
fallenden Bezugssysteme sind lokal hinsichtlich 
Orientierung der Achsen und Geschwindigkeit physikalisch 
gleichberechtigt. G"abe es eine das ganze Universum 
erf"ullende Substanz, wie den hypothetischen "Ather 
"alterer Theorien, der in jedem Raumpunkt das Bezugssystem
bevorzugt in dem die "Atherteilchen ruhen, dann w"urde 
man erwarten, dass sich eine lokale Bewegung relativ 
zu dieser Substanz auf das physikalische Geschehen auch 
lokal auswirkt. Ein klassisches Experiment ist das 
von Albert Michelson (1852-1931) und Edward Morley,
(1838-1923), das zun"achst 1881 in Potsdam von Michelson 
alleine und dann 1887 von beiden mit einer wesentlich 
verbesserten Apparatur und einer verzehnfachten Genauigkeit 
an der Universit"at von Cleveland Ohio (USA) wiederholt wurde. 
Dieses Experiment besteht aus einem nach Michelson benannten 
Interferometer, mit dessen Hilfe m"ogliche 
Richtungsabh"angigkeiten der zwischen Hin- und R"uckweg 
gemittelten Lichtgeschwindigkeit gemessen werden k"onnen.
Die Ergebnisse waren negativ -- und blieben es seither. 
Ein positiver Effekt w"urde sich beispielsweise in einer 
richtungsabh"angigen Variation $\Delta c$ der Lichtgeschwindigkeit $c$ 
manifestieren, die mittlerweile durch modernste Experimente 
(mit rotierenden optischen Resonatoren) von Sven Herrmann und 
Mitarbeitern (ZARM Bremen, Humboldt Universit"at Berlin) aus 
dem Jahr 2009 eingeschr"ankt sind auf~\cite{Herrmann.etal:2009}  
\begin{equation}
\label{eq:MichelsonMorleyGenauigkeit}  
\frac{\Delta c}{c}<10^{-17}\,.
\end{equation}
\item[3.]
\textbf{Die Universalit"at der gravitativen Rotverschiebung (UGR):}
Standarduhren werden durch Gravitationsfelder in einem 
zweifachen Sinn universell beeinflusst. Erstens: 
Verschiedene Uhren (etwa eine C"asium-Atomuhr und eine 
Uhr basierend auf einem Wasserstoff-Maser) die sich im 
Gravitationsfeld auf der gleichen Bahn bewegen (etwa beide 
an Bord der ISS) gehen synchron. Zweitens: Uhren gleicher 
Bauart die sich im Gravitationsfeld auf verschiedenen 
Bahnen bewegen  und deren relativer Gang verglichen wird, 
etwa durch den Austausch elektromagnetischer Signale, 
zeigen immer den gleichen universellen Gangunterschied, 
der bei schwachen Gravitationsfeldern durch die Formel
\begin{equation}
\label{eq:Rotverschiebung}
\frac{\Delta\nu}{\nu}=(1+\alpha)\frac{\Delta U}{c^2}
\end{equation} 
ausgedr"uckt wird. Dabei ist $\Delta\nu$ die Differenz der 
Taktfrequenzen beider Uhren und $\nu$ ihr Mittel. $\Delta U$ 
bezeichnet die entsprechende Differenz der Gravitationspotentiale 
auf denen sich die Uhren befinden und $c$ ist die bereits 
erw"ahnte Naturkonstante (Lichtgeschwindigkeit). Der zus"atzlich 
auftretende Parameter $\alpha$ ist gem"a"s der Allgemeinen
Relativit"atstheorie Null zu setzen. Sollte es jedoch eine
Abweichung von UGR geben, so w"are $\alpha$ von Null 
verschieden, und zwar in einer sowohl von der Art der 
verwendeten Uhr als auch von der Lokalisation in Raum und 
Zeit anh"angigen Weise. Erstaunlicherweise liegt der 
bisher beste Test von UGR mehr als 36 Jahre zur"uck. 
In diesem wurde eine Wasserstoff-Maser Uhr mit einer 
Rakete auf eine H"ohe von mehr als zehntausend Kilometer 
bef"ordert und w"ahrend der fast zweist"undigen Flugzeit 
durch Mikrowellensignale mit einer Wasserstoff-Maser Uhr
auf der Erdoberfl"ache verglichen. Dieses Experiment, 
das zusammen vom \emph{Smithsonian Astrophysical Observatory}
in Cambridge Massachusetts und dem \emph{George C.~Marshall 
Space Flight Center} in Huntsville Alabama durchgef"uhrt 
wurde und als \emph{Gravity-Probe-A} in die Geschichte 
einging, ergab folgende obere Schranke an $\alpha$: 
\begin{equation}
\label{eq:AlphaObSchrankeGPA}
\alpha\leq 7\times 10^{-5}\,. 
\end{equation}
Auf der Erdoberfl"ache bedingt eine vertikale Verschiebung 
nach oben eine Zunahme des Gravitationspotentials und 
damit der Frequenz einer Uhr. Pro Meter betr"agt diese 
relative Frequenzerh"ohung etwa $10^{-16}$. Die zur Zeit 
genauesten Uhren sind so genannte optische Atomuhren (s.u.). 
Diese weisen relative Ganggenauigkeiten von fast $10^{-17}$
auf, dass hei"st, sie gehen in $10^{17}$ Sekunden, das sind 
etwa 3,2 Milliarden Jahre, nur um eine Sekunde vor oder nach. 
Damit wurde im Jahre 2010 am National Institute of Standards 
and Technology (NIST) in Boulder tats"achlich die Rotverschiebung 
"uber eine Distanz von nur $33$ Zentimeter 
nachgewiesen~\cite{Chou.EtAl:2010}. 
\end{itemize}

Bemerkenswert an dieser Dreiteilung des Einstein'schen 
"Aquivalenzprinzips ist die Diskrepanz zwischen der hohen 
Genauigkeit mit der UFF und LLI "uberpr"uft wurden und der 
dazu deutlich abgeschlagenen Genauigkeit unserer Kenntnis 
von UGR. W"urde man letztere wesentlich steigern k"onnen, 
h"atte man damit auch eine wesentliche Verbesserung in unserer 
Kenntnis des gesamten "Aquivalenzprinzips erreicht, das ja nur 
als so gut best"atigt gelten kann wie sein am schlechtesten 
bekannter Teil. "Uberraschenderweise sind in letzter Zeit 
diesbez"uglich Hoffnungen geweckt worden, dass die Quantenmechanik 
hier Abhilfe schaffen kann. Doch bevor wir darauf n"aher eingehen, 
m"ussen wir uns kurz der wichtigen Frage zuwenden, welche 
fundamentaleren physikalischen Einsichten "uberhaupt von einer 
solchen Steigerung erhofft werden d"urfen.  
  
\section{Die Niederenergie-Front der Teilchenphysik}
Die moderne Teilchenphysik geht davon aus, dass es genau vier 
fundamentale Wechselwirkungen gibt, n"amlich die nur auf 
subatomaren Distanzen relevanten \emph{starke} und 
\emph{schwache Wechselwirkung}, sowie den \emph{Elektromagnetismus}
und die \emph{Gravitation}. Letztere sind langreichweitig und 
bestimmen die physikalischen Strukturen von atomaren L"angenskalen 
bis hinauf zu den kosmologischen. Doch welche Wechselwirkung   
rechnen wir eigentlich zur Gravitation? Angenommen es g"abe 
in der Natur neben den uns bereits bekannten Feldern noch 
weitere, die nur sehr schwache Kopplungen an die uns gegebenen 
Materieformen besitzen, so dass die mikroskopische Struktur der 
Materie davon nicht wesentlich beeinflusst w"urde. K"onnten wir 
von solchen Feldern "uberhaupt Kenntnis~haben?

\begin{wrapfigure}[22]{r}{0.5\linewidth}
\centering\includegraphics[scale=0.65]{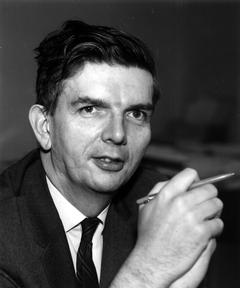}
\caption{{\footnotesize Robert Henry Dicke (1917-1997) ist einer 
der Pioniere f"ur die Konzeption und Durchf"uhrung von 
Pr"azisionsexperimenten zum Test der ART, die diese bisher 
s"amtlich mit Bravour bestanden hat.}}
\end{wrapfigure}
Diese Frage hat den amerikanischen Atom-, Laser- und Astrophysiker 
Robert Henry Dicke von der Universit"at Princeton seit Ende der 
50er Jahre intensiv besch"aftigt. Er wies wiederholt darauf hin, 
dass sich unter Umst"anden solche Felder durch charakteristische 
Modifikationen des Gravitationsgesetzes bemerkbar machen k"onnten 
die dem "Aquivalenzprinzip widersprechen.
Dabei kommen solche Felder in 
Betracht, bei denen das Austauschteilchen keinen Spin (inneren Drehimpuls) 
besitzt, so genannte Skalarfelder, oder einen Spin von Eins (in 
Einheiten von $\hbar$), dann spricht man von Vektorfeldern. 
Skalarfelder f"uhren zu einer zus"atzlichen Anziehung, Vektorfelder zu 
einer Absto"sung. Au"serdem muss noch danach unterschieden werden, ob das 
Feld langreichweitig ist, was der Fall ist, wenn das Austauschteilchen 
masselos ist, oder ob das Austauschteilchen eine Masse besitzt,
in welchem Fall die zus"atzliche Kraft nochmals exponentiell mit 
dem Abstand unterdr"uckt ist. Zum Beispiel hat das Austauschteilchen 
der elektromagnetischen Wechselwirkung, das Photon, keine Masse und 
Spin 1, ist also 
langreichweitig (Potential f"allt umgekehrt proportional zum Abstand 
aber nicht exponentiell ab) und absto"send (gleiche Ladungen sto"sen 
sich ab). Die Modifikation des Gravitationsgesetzes, die 
beispielsweise durch ein masseloses Skalarfeld hervorgerufen w"urde, 
ist in Anhang~2 n"aher dargestellt. 
Dabei ist der tiefere Grund f"ur die Verletzung des "Aquivalenzprinzips 
darin zu sehen, dass die Kopplung dieses Zusatzfeldes nicht an die totale 
Masse erfolgt, wie in der ART, also die Masse im Sinne von $E=mc²$, zu der 
ausnahmslos \emph{alle} Energieformen beitragen, sondern an eine "`Ladung"' 
anderer Art, etwa der Baryonenzahl, der Leptonenzahl, oder einer Kombination 
aus beiden. Als Konsequenz ergibt sich ein von der Komposition der Materie 
abh"angiges Verh"altnis von schwerer zu tr"ager Masse und damit eine 
Verletzung der UFF.

Nun ist die Baryonenzahl eines Atoms gleich der Atomzahl $A$, also 
der Summe aus der Ladungszahl $Z$ (Anzahl der Protonen) und der 
Anzahl $N$ der Neutronen. Die Leptonenzahl ist gleich der Anzahl der 
Elektronen und damit gleich $Z$, vorausgesetzt das Atom ist neutral.
Welche Kombination dieser Gr"o"sen in der Zusatzkraft eine Rolle spielt 
ist theorieabh"angig. Will man sich hier nicht festlegen, so ist es 
experimentell wichtig, verschiedene Materialien zu betrachten und 
buchst"ablich gegeneinander abzuw"agen, um so ein m"oglichst breites 
Spektrum an Ladungstypen zu erfassen. Dabei ist der Fall der 
Kopplung an die Baryonenzahl besonders schwierig, da das Verh"altnis 
der Baryonenzahl zur Gesamtmasse von Atom zu Atom nur aufgrund der 
Bindungsenergie variiert, die nur im Promillebereich zur Gesamtmasse 
beitr"agt, w"ahrend die Variationen etwa der Neutronenzahl im Prozentbereich 
liegen. Ein Elementenpaar, in dem die Variation der Baryonenzahl pro 
Gesamtmasse vergleichsweise gro"s ist, ist Beryllium-Titan. Dies ist einer
der Gr"unde, weshalb gerade diese Materialpaarung im oben 
zitierten Torsionswaagenexperiment der "`E"ot-Wash"'-Gruppe verwendet wurde.

\begin{wrapfigure}[22]{r}{0.5\linewidth}
\centering\includegraphics[scale=0.45]{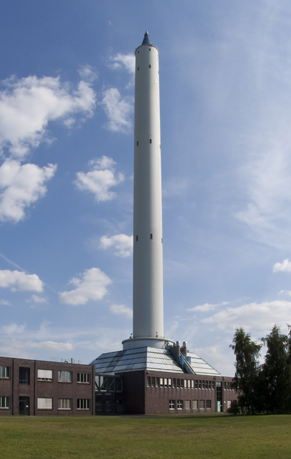}
\caption{\small Der 146 Meter hohe Fallturm in Bremen, in dem 
zur Zeit vorbereitende Experimente zum "Aquivalenzprinzip 
an frei fallenden Quantengasen durchgef"uhrt werden.\medskip {}}
\end{wrapfigure}
W"ahrend bisher die zus"atzlichen, langreichweitigen Felder 
rein hypothetischer Natur waren, treten diese in vereinheitlichten 
Theorien nat"urlich auf. Als erstes Beispiel seien verallgemeinerte 
Gravitationstheorien genannt, in denen die Gravitationskonstante 
tats"achlich nicht konstant sondern gleich einem solchen Feld ist. 
Diese Theorien werden oft nach Carl Brans (geb. 1935) und 
Robert Dicke  (1916-1997) einfach ``Brans-Dicke-Theorien'' genannt,
wurden aber zuvor bereits von Pascual Jordan (1902-1980) und 
Markus Fierz (1912-2006) untersucht.  Typischerweise treten 
langreichweitige Skalarfelder in Theorien auf, die auf Raum-Zeiten 
mit mehr als 4 Dimensionen beruhen. In diese Familie geh"oren so 
genannte Kaluza-Klein Theorien und auch die String-Theorie, 
in der das so genannte Dilaton-Feld die Rolle des skalaren 
Feldes spielt. Bereits 1994 wiesen Thibault Damour vom IHES 
(Institut des Hautes \'Etudes Scientifiques) in Bures-sur-Yvette 
bei Paris und Alexander Polyakov von der Universit"at Princeton 
in einer gemeinsamen Publikation \cite{Damour.Polyakov:1994a} 
darauf hin, dass in der Stringtheorie in dieser Weise 
UFF-Verletzungen ab Gr"o"senordnungen $10^{-15}$ erwartet werden 
k"onnen. Dieser Bereich wird aber gerade durch die in n"achster 
Zukunft geplanten Experimente erstmals betreten werden!

Der Mechanismus der UFF-Verletzung, der durch die Stringtheorie nahe 
gelegt wird, ist selbst von sehr allgemeiner Art. In der Stringtheorie 
sind n"amlich die Kopplungsst"arken der verschiedenen Wechselwirkungen 
keine wirklichen Konstanten mehr, sondern abh"angig vom Dilaton-Feld, 
dessen Werte raumzeitlich variieren k"onnen. Damit k"onnen aber 
auch die Teilchenmassen raumzeitlich variieren, denn diese h"angen von 
den Kopplungsst"arken der Wechselwirkungen ab. Dann ergibt sich aber 
ein nicht-verschwindender E"otv"os Parameter f"ur Teilchenpaare immer 
dann, wenn die Variation der St"arke einer der Kopplungen von denen
die Teilchenmassen abh"angen \emph{unterschiedliche} relative 
Beitr"age zu den beiden Teilchenmassen liefert. Diese Idee von 
Damour und Polyakov f"uhrt zu m"oglichen UFF-Verletzungen, die 
sich in das allgemeine oben beschriebene Schema einordnen und 
die im Falle eines Atoms mit den verschiedenen Energie-Beitr"agen 
verbunden sind. Man erh"alt so zwei wesentliche Beitr"age zum 
E"otv"os-Faktor, die jeweils aus der Differenz von zwei Termen 
bestehen, die unterschiedliche Abh"angigkeiten von den 
elementaren Ladungen besitzen. Ein Term ist proportional zu 
$A^{-1/3}$, wobei $A$ die Atomzahl ist, der andere zu $Z^2A^{-4/3}$,
wobei $Z$ die Ladungszahl (Protonenzahl) ist. 
Will man experimentell \emph{diese} M"oglichkeit der 
UFF-Verletzung testen, muss man darauf achten, solche 
Materialpaarungen zu w"ahlen, f"ur die die genannten 
Terme m"oglichst stark voneinander differieren.
Diese Vorgabe steht oft im Kontrast zu anderen Forderungen, 
die u.U. das Experiment "uberhaupt erst m"oglich machen.
In diesem Fall m"ussen Kompromissl"osungen gefunden werden.

Beispielsweise sind die relevanten Beitr"age 
f"ur das Paar Titan-Platin bis zu einen Faktor 50 gr"o"ser 
als f"ur das Paar der Rubidium Isotope ${}^{85}\!Rb$--${}^{87}\!Rb$. 
Letztere eignen sich aber f"ur atominterferometrische
Experimente und wurden tats"achlich benutzt, um die 
Schwerebeschleunigung des Gravitationsfeldes der Erde zu 
messen, gewisserma"sen also als quantenmechanisches 
Gravimeter. Erste Pionier-Experimente ergaben eine 
obere Schranke f"ur den E"otv"os Faktor dieses 
Isotopenpaares von etwas "uber $10^{-7}$~\cite{Fray.Weitz:2009}. 
Weitere Experimente sind geplant oder werden 
gegenw"artig durchgef"uhrt~\cite{vanZoest:2010}, so am 
Bremer Fallturm innerhalb des Projekts QUANTUS (Quantengase 
unter Schwerelosigkeit). Ebenso sind Erweiterungen auf g"unstigere 
Materialpaarungen geplant.

\section{"Aquivalenzprinzip und Quantenmechanik}
Die oben gegebene Formulierung des "Aquivalenzprinzips benutzt 
wesentlich die Konzepte der klassischen Physik, insbesondere 
das des \emph{Testteilchens} und der \emph{Teilchenbahn}. Eine
w"ortliche "Ubertragung in die Quantenmechanik ist deshalb
nicht m"oglich. Bisweilen wurde sogar die Unm"oglichkeit einer 
"Ubertragung behauptet, da die Quantenmechanik dem "Aquivalenzprinzip 
angeblich "`offensichtlich"' widerspreche und "`den rein klassischen 
geometrischen Standpunkt unterh"ohle"' (vgl. \cite{Greenberger.Overhauser:1980}).
Als Standardbeispiel kann das klassische COW-Experiment von Roberto Collela, 
Albert Obverhauser (beide von der Purdue University, Indiana USA) und 
Samuel Werner (Ford Motor Company) dienen, das diese 1975 am  
Ford Nuclear Reactor der Universit"at von Michigan in Ann Arbor 
durchf"uhrten; siehe~\cite{Greenberger.Overhauser:1980}. Dort wurden 
Interferenzen von Materiewellen kalter, im Gravitationsfeld der Erde 
frei fallender Neutronen untersucht und erstmals deren gravitationsfeldbedingte 
Phasenverschiebungen direkt gemessen. Diese h"angen unmittelbar von der 
tr"agen Masse der Neutronen ab,  einfach deshalb, weil die L"ange der 
Materiewelle, die das Beugungsmuster bestimmt, umgekehrt proportional 
zum Impuls des Teilchens ist (die Beugung geschieht im COW-Experiment 
an einem Silizium-Kristall). Somit h"angt auch das Beugungsmuster von 
der tr"agen Masse ab. Dies -- so die weit verbreitete 
Behauptung~\cite{Greenberger.Overhauser:1980} -- widerspreche dem 
"Aquivalenzprinzip, gem"a"s dem die Bahnen von Teilchen im Gravitationsfeld 
nicht von deren Masse abh"angen d"urfen. 

Wie man sich aber leicht klar macht, ist diese Begr"undung nicht 
stichhaltig. Einerseits ist die Bewegung wegen der erzwungenen Beugung
nicht frei. Andererseits besagt  das "Aquivalenzprinzip \emph{nicht} die 
Unmessbarkeit der tr"agen oder schweren Masse, sondern deren universelle 
Gleichheit. Die Beugung entspricht einer Impulsmessung, die nicht anders 
als in der klassischen Mechanik R"uckschl"usse  auf die tr"age Masse 
erlaubt; und die schwere Masse ermittelt man durch einfaches Wiegen! 
Man kann leicht andere Beispiele dieser Art konstruieren. 
So sind die quantenmechanischen Energieniveaus von Neutronen, 
die sich oberhalb einer ideal reflektierenden horizontalen 
Ebene im Gravitationsfeld der Erde bewegen, proportional zur
dritten Wurzel des Verh"altnisses zwischen dem Quadrat der schweren 
Masse und der tr"agen Masse. Diese Energieniveaus sind trotz der winzigen 
bindenden Wirkung des Gravitationsfeldes (die Energieniveaus sind 
im Bereich einiger Pico-Elektronenvolt!) an ultrakalten Neutronen 
am Institut Laue-Langevin in Grenoble tats"achlich gemessen 
worden~\cite{Nesvizhevsky:Nature-2002}. 

Die M"oglichkeit der Messung von Gr"o"sen mit verschiedenen 
Abh"angigkeiten von der tr"agen und der schweren Masse ist kein 
spezifisch quantenmechanischer Umstand und widerspricht in 
keiner Weise dem "Aquivalenzprinzip -- solange sich die beiden 
Massen als gleich herausstellen! Viel interessanter ist die 
umgekehrte Bemerkung, dass man dadurch, dass diese Gr"o"sen 
durch die hohe Genauigkeit mit der quantenmechanische Messungen 
oft m"oglich sind, verbesserte Tests des "Aquivalenzprinzips 
erm"oglichen. Damit kehren wir auf den Weg zur"uck,
den wir am Ende des Abschnitts "uber das "Aquivalenzprinzip 
vor"ubergehend verlassen haben. 

Die vielleicht naivste Art die Gravitationsbeschleunigung 
auf der Oberfl"ache der Erde zu messen ist die, einen 
Gegenstand einfach fallen zu lassen und die durchfallene H"ohe 
nach einer festen Zeit nachzumessen. Macht man dies mit 
einzelnen Atomen, so kann man sich den Umstand zu Nutze 
machen, dass der Schwerpunktsbewegung des Atoms gem"a"s der 
Quantenmechanik eine eigene Materiewelle zugeordnet wird, 
die durch geeignete Vorrichtungen koh"arent geteilt und
wieder zusammengef"uhrt werden kann. Beobachtungen der 
Interferenz geben dann sehr pr"azise Auskunft "uber die 
Ausbreitung der Welle im Gravitationsfeld. Eine solche 
quantenmechanische Messung der Schwerebeschleunigung eines 
Atoms im Gravitationsfeld der Erde wurde 1999 von Achim Peters, 
Keng Yeow Chung und Steven Chu von der Universit"at Stanford 
in Kalifornien ausgef"uhrt~\cite{Peters.Chung.Chu:1999}. 
Die erreichte Genauigkeit lag bei drei Teilen eines Milliardstels 
($3\times 10^{-9}$). Diesen Wert verglichen sie mit dem Wert der 
Schwerebeschleunigung eines frei fallenden makroskopischen 
K"orpers aus Glas, dessen Fall sie im gleichen Labor mit 
Hilfe von Lasern ebenfalls sehr genau gemessen hatten. 
Innerhalb der Messgenauigkeit ergaben sich keine 
Unterschiede, so dass man ihr Experiment durch die Aussage 
zusammenfassen kann, dass der E"otv"os-Faktor f"ur 
einzelne C"asium-Atome und Glas kleiner als $10^{-9}$ ist:
\begin{equation}
\label{eq:Eoetvos_Cs-Glas}
\eta(\text{Cs},\text{Glas})<10^{-9}\,.
\end{equation} 
Das mag angesichts des um vier Gr"o"senordnungen besseren Wertes den 
man mit Torsionswaagen erhalten hat nicht sonderlich beeindrucken, 
ist aber doch in folgender Hinsicht bemerkenswert: Die Materialpaarung 
ist hier zwischen einzelnen Atomen einerseits, die sich dar"uber hinaus 
noch in verschr"ankten Zust"anden (hinsichtlich der Lokalisation 
des Schwerpunktes) befinden, und makroskopischer Materie in 
unverschr"ankten Zust"anden. Damit ist ein direkter Test der 
G"ultigkeit von UFF zwischen klassischer und "`Quantenmaterie"' gelungen. 

"Uberraschenderweise soll dieses Experiment nachtr"aglich 
aber auch als Test von UGR verstanden werden, wie einige 
der Autoren zusammen mit anderen mehr als 10 Jahre 
sp"ater (Februar 2010) in einem viel diskutierten Artikel 
in der renommierten Englischen Zeitschrift \emph{Nature} 
argumentieren~\cite{Mueller.etal:2010}. W"are dies 
richtig, so k"ame dies einer Sensation gleich, denn w"ahrend 
eine obere Schranke von $10^{-9}$ um vier Gr"o"senordnungen
"uber der bereits bekannten Schranke an m"ogliche 
UFF-Verletzungen liegt, ist sie doch gleichzeitig um vier 
Gr"o"senordnungen kleiner (also besser) als bisherige 
Schranken an m"ogliche UGR-Verletzungen. Damit w"urde 
auch das gesamte Einstein'sche "Aquivalenzprinzip eine 
um vier Gr"o"senordnungen bessere Best"atigung erfahren!

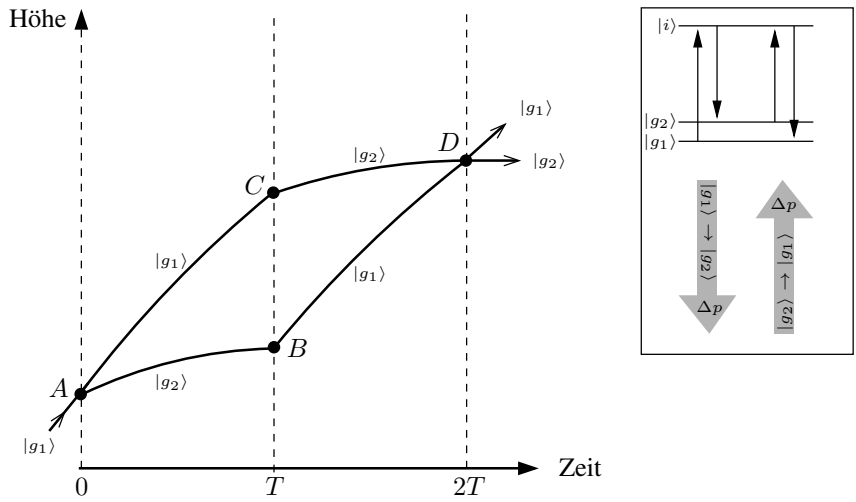
\begin{figure}
\centering\includegraphics[width=0.814\linewidth]{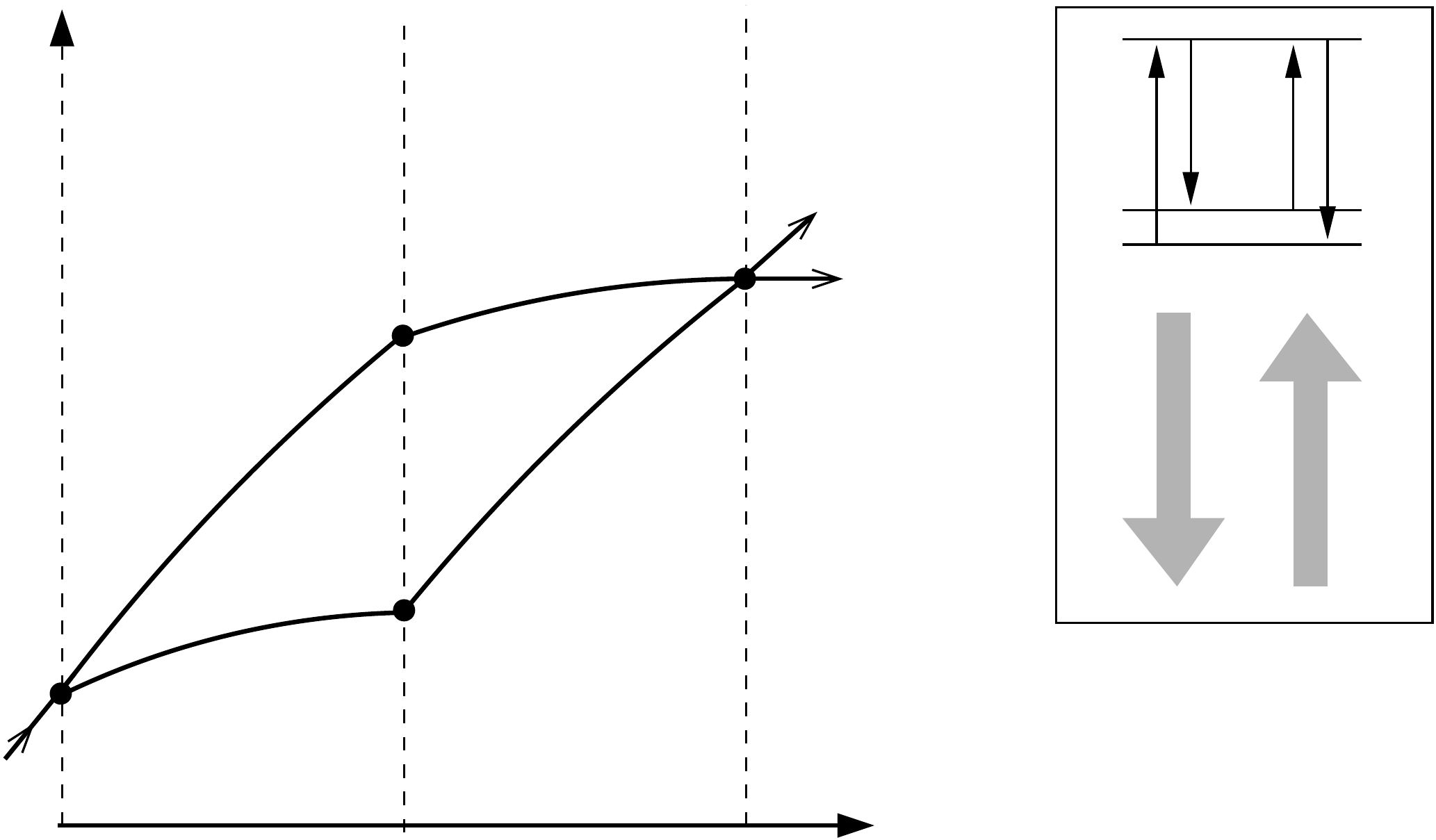}
\put(-80,125){\tiny $\vert g_1\rangle$}
\put(-80,133){\tiny $\vert g_2\rangle$}
\put(-75,169){\tiny $\vert i\rangle$}
\put(-127,138){\tiny $\vert g_1\rangle$}
\put(-122,118){\tiny $\vert g_2\rangle$}
\put(-190,120){\tiny $\vert g_2\rangle$}
\put(-190,75){\tiny $\vert g_1\rangle$}
\put(-265,34){\tiny $\vert g_2\rangle$}
\put(-315,10){\tiny $\vert g_1\rangle$}
\put(-265,80){\tiny $\vert g_1\rangle$}
\put(-305,30){\small $A$}
\put(-215,45){\small $B$}
\put(-231,107){\small $C$}
\put(-158,123){\small $D$}
\put(-320,170){\small H"ohe}
\put(-112,0){\small Zeit}
\put(-295,-7){\small $0$}
\put(-223,-7){\small $T$}
\put(-152,-7){\small $2T$}
\put(-60,62){\tiny $\Delta p$}
\put(-32,101){\tiny $\Delta p$}
\put(-57,111){\tiny\begin{rotate}{-90}%
$\vert g_1\rangle\rightarrow\vert g_2\rangle$\end{rotate}}
\put(-25,55){\tiny\begin{rotate}{90}%
$\vert g_2\rangle\rightarrow\vert g_1\rangle$\end{rotate}}
\caption{\label{fig:FigKasevichChu}\small
Schematischer Verlauf der Atomstrahlen im Kasevich-Chu-Interferometer.
Unter Einfluss des vertikal nach unten gerichteten Gravitationsfeldes 
der Erde sind diese gekr"ummt. Gezeigt ist die vertikale H"ohe $h$ als 
Funktion der Zeit. Durch zwei 
gegenl"aufige Laserpulse in vertikaler Richtung wird im weiteren Verlauf 
jedes Atom mit einer gewissen Wahrscheinlichkeit entweder vom wahren 
Grundzustand $\vert g_1\rangle$ in den hyperfein gehobenen Grundzustand 
$\vert g_2\rangle$ gebracht und dabei der feste Impulsbetrag $\Delta p$ 
nach unten (in Richtung des Gravitationsfeldes) "ubertragen, oder umgekehrt. 
Anfangs sind alle Atome im wahren Grundzustand $\vert g_1\rangle$ und 
bewegen sich vertikal nach oben (entgegen der Richtung des 
Gravitationsfeldes). Bei $A$ wird der "Ubergang 
$\vert g_1\rangle\rightarrow\vert g_2\rangle$ mit 50\% Wahrscheinlichkeit 
getriggert, so dass Strahlteilung erfolgt. Bei $B$ und $C$ werden die 
jeweiligen Zust"ande mit 100\% Wahrscheinlichkeit vertauscht, und bei $D$ 
wieder mit 50\% Wahrscheinlichkeit aufgespalten. Die Interferenzmessung 
erfolgt dann an einem der beiden Teilstrahlen rechts von $D$, die 
jeweils aus einer Superposition von Atomen entlang der Wege $ABD$
und $ACD$  bestehen. Zwischen den Pulsen bei $A$ und $B,C$, bzw. 
letzteren und $D$ besteht der zeitliche Abstand $T$, der im 
tats"achlich ausgef"uhrten Experiment von Chung et. al. 
160 Millisekunden betrug. Die resultierende H"ohendifferenz zwischen 
$C$ und $B$ war nur 0,12\,mm.}
\end{figure}

Die zu diesem Schluss notwendige Interpretation des physikalischen 
Geschehens ist aber bis heute sehr umstritten. In "uberraschender 
Weise r"uhrt der kritische Punkt dabei an fundamentalen Fragestellungen, 
die sowohl unser Verst"andnis der Quantenmechanik betreffen, als auch die 
Antwort auf die Frage: "`Was ist eine Uhr?"', die zu den Grundlagen der 
ART geh"ort. Um einigerma"sen nachvollziehen zu k"onnen, worum es in 
dieser Auseinandersetzung geht, m"ussen wir etwas genauer auf 
das Experiment von 1999 eingehen. Bei diesem handelt es 
sich um ein so genanntes Kasevich-Chu-Interferometer, bei dem 
die Teilung und Reflexion von Materiewellen durch 
laserinduzierte "Uberg"ange zwischen sehr dicht beieinander 
liegenden Grundzustandsniveaus (Hyperfeinaufspaltung) realisiert 
werden; siehe Abbildung\,\ref{fig:FigKasevichChu}.
Bestimmt man die Phasenverschiebung in Kasevich-Chu-Interferometer theoretisch 
mit Hilfe der Quantenmechanik (was mit Hilfe der Pfadintegralmethode sehr 
"ubersichtlich gelingt), so erh"alt man das Produkt aus 
Gravitationsbeschleunigung $g$, dem Impuls"ubertrag dividiert durch das 
reduzierte Planck'sche Wirkungsquantum $\Delta p/\hbar$ (wobei $\hbar=h/2\pi$) 
und dem Quadrat $T^2$ der Zeit zwischen den Laserpulsen; also 
\begin{equation}
\label{eq:DeltaPhi}
\Delta\Phi =g\ T^2\ \frac{\Delta p}{\hbar}\,.
\end{equation}
"Uber die Richtigkeit dieser Formel besteht Einigkeit, nicht jedoch 
"uber ihre Interpretation als Rotverschiebung. (Eine ausf"uhrliche
Diskussion mit allen Herleitungen findet man in \cite{Giulini:2012}.)
Zu dieser Interpretation gelangt man so: Angenommen jedes Atom sei eine Uhr, 
die mit der Frequenz $\nu$ tickt. Nach der ART ticken die "`Uhren"' 
im oberen Teilstrahl, der sich in einem um den Betrag $\Delta U$ 
h"oheren Gravitationspotential bewegen, mit einer um 
$\Delta\nu=\nu\Delta U/c^2$ h"oheren  Frequenz. Nach einer 
Zeitspanne $T$ haben also die "`Uhren"' des oberen 
Teilstrahls eine um $T\Delta\nu$ h"ohere Anzahl von 
Schl"agen absolviert, was einer Phasenverschiebung $\Delta\Psi$ 
von $2\pi$ mal dieser Anzahl entspricht. Ber"ucksichtigt man 
jetzt noch, dass der Unterschied $\Delta U$ des Gravitationspotentials 
einfach das Produkt des H"ohenunterschieds $\Delta h$ der 
Teilstrahlen mit der "ortlichen Gravitationsbeschleunigung $g$ ist, 
und dass der H"ohenunterschied $\Delta h$ seinerseits gleich ist 
dem Produkt des Geschwindigkeitsunterschieds 
$\Delta v=\Delta p/m$ ($m$ ist die tr"age Masse) mit der 
Zeitspanne $T$, so ergeben die einzelnen Schritte zusammen
folgendes Resultat:
\begin{equation}
\label{eq:DeltaPhiTheo}
\begin{split}
\Delta\Psi 
&=2\pi\Delta\nu T
=2\pi\nu\frac{\Delta U}{c^2} T
=2\pi\nu\frac{g\Delta h}{c^2} T
=2\pi\nu\frac{g\Delta p}{mc^2}\ T^2\\
&=\left(\frac{\nu}{mc^2/h}\right) g\ T^2\ \frac{\Delta p}{\hbar}\,.
\end{split}
\end{equation}
Machen wir noch von der Definition der Compton-Frequenz Gebrauch, 
\begin{equation}
\label{eq:DefComptonFrequenz}
\nu_c=mc^2/h\,,
\end{equation}
dann sehen wir durch Vergleich von (\ref{eq:DeltaPhiTheo}) mit 
(\ref{eq:DeltaPhi}), dass eine Gleichheit $\Delta\Psi=\Delta\Phi$ 
genau dann besteht, wenn $\nu=\nu_c$. Wenn man also annimmt, 
dass Atom sei per se eine Uhr mit der Taktfrequenz $\nu=\nu_c$, 
dann entspricht die Phasenverschiebung  (\ref{eq:DeltaPhi}) der 
entsprechenden Rotverschiebung dieser Uhr. Die logische Umkehrung 
folgt nat"urlich nicht, so dass aus der Korrektheit von 
(\ref{eq:DeltaPhi}) nicht auf das Zutreffen dieser Interpretation 
als Rotverschiebung geschlossen werden kann. 

F"ur ein C"asium-Atom ist die Compton-Frequenz enorm hoch, 
n"amlich $3\times 10^{25}Hz$! 
In der Tat muss man schon ganz besondere Uhren haben, um beim hier 
diskutierten Experiment "uberhaupt eine Rotverschiebung zu messen, 
denn die beiden Teilstrahlen der Atome haben eine vertikale 
Verschiebung im Gravitationsfeld von nur $0.12\,mm$. 
Dazu ben"otigt man Uhren die dreitausend mal genauer gehen 
als die beste heute bekannte Uhr (optische Atomuhr), die, wie oben 
gesehen, eine Rotverschiebung gerade "uber 33~cm nachweisen kann. 
Soll dar"uber hinaus die relative Genauigkeit dieser Messung 
sogar $10^{-9}$ betragen, wie in der angegebenen Interpretation 
behauptet wird, so muss diese Uhr sogar um den Faktor $10^{12}$ 
genauer gehen als die beste Atomuhr. Diesem Faktor entspricht 
in etwa die Steigerung von den typischen Frequenzen einer Atomuhr 
zur Compton-Frequenz der hypothetischen "`Compton Uhr"'. 
%Folgt man momentan dieser Interpretation, so ergibt
%das Experiment f"ur die Rotverschiebung pro Meter im Gravitationsfeld 
%der Erde am Ort des Experiments den Wert (zum Vergleich mit der 
%theoretischen Voraussage): 
%
%\begin{alignat*}{2}
%\label{eq:RotverschiebungProMeter}
%&\zeta_{exp}&&=(1,090\,322\,683\pm 0.000\,000\,003)\times 10^{-16}\,,\\ 
%&\zeta_{theo}&&=(1,090\,322\,675\pm 0.000\,000\,006)\times 10^{-16}\,.
%\end{alignat*}

Eine offensichtliche Kritik an dieser Interpretation ist diese: 
Im Interferometer sind die C"asium-Atome fast immer in 
Energie-Eigenzust"anden. Nur w"ahrend der sehr kurzen 
Wechselwirkungsphase mit dem Laser springen manche der Atome 
von einem Eigenzustand in einen nahe benachbarten. 
Energieeigenzust"ande sind aber station"ar, also zeitlich 
unver"anderlich. Ein wirkliche Uhr, von der man die Zeit auch 
tats"achlich ablesen kann, wird aber niemals in einem station"aren 
Zustand sein, denn eine solche Uhr muss ihren Zustand mit der Zeit 
ver"andern: Ihr "`Zeiger"' muss sich bewegen! 
Deswegen sind die Atome in herk"ommlichen Atomuhren auch immer 
in Superpositionen zweier Energie-Eigenzust"ande. Die \emph{Differenz} 
der zugeh"origen Energien geteilt durch das Planck'sche Wirkungsquantum 
gibt dann die Frequenz an, mit der die Uhr tickt. Bei den bisher 
am schnellsten tickenden Uhren, den  optischen Atomuhren, ist 
diese Frequenz in der Gr"o"senordnung der Frequenz sichtbaren 
Lichts (deshalb das Adjektiv "`optisch"'). Wie wir gesehen 
haben, ist dies um viele (s.o.) Gr"o"senordnungen geringer 
als die Compton-Frequenz. 

Diese etwas seltsame Vorstellung, dass jedem St"uck Masse bzw. 
Energie von sich aus bereits eine Art "`Eigenuhr"' zugeordnet werden 
kann, findet sich bereits in den ersten Arbeiten Louis de Broglies,
dem Urheber der Idee der Materiewelle. In seiner Doktorarbeit 
"`Recherche sur la Th\'eorie des Quanta"' aus dem Jahr 1924 umschrieb 
er seine Hypothese (in deutscher "Ubersetzung) so:

\begin{wrapfigure}[22]{r}{0.40\linewidth}
\centering\includegraphics[scale=0.38]{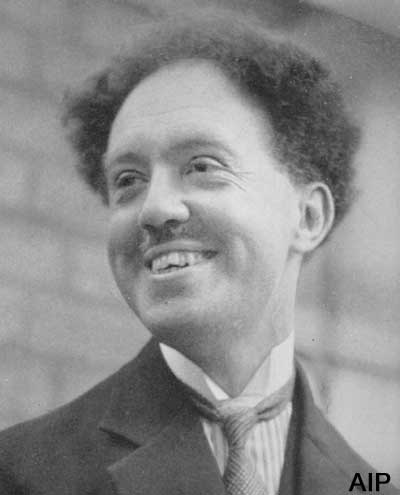}
\caption{\small Louis de\,Broglie (1892-1987) konzipierte 1923 im 
Rahmen seiner Doktorarbeit die Vorstellung der Materiewellen,
wof"ur er 1929 den Nobelpreis f"ur Physik erhielt.}
\end{wrapfigure}
\noindent \textit{"`Die Grundidee der Quantenmechanik ist wohl die Unm"oglichkeit, ein isoliertes Energiest"uck zu 
betrachten, ohne diesem eine gewisse Frequenz zuzuordnen.[...]
Man darf daher annehmen, dass zufolge einem fundamentalen Naturgesetz (une grande loi de la Nature) 
mit jedem Energiest"uck von der Eigenmasse $m_0$ ein periodisches 
Ph"anomen von der Frequenz $\nu_0$ nach der Beziehung [angegeben 
wird dann die Compton Frequenz] verbunden ist. [...]
Diese Hypothese ist die Grundlage unserer Theorie: Sie gilt, wie alle 
Hypothesen, so weit, wie die aus ihr zu ziehenden Folgerungen"'}

Man beachte, dass die Natur des "`periodischen Ph"anomens"'
nicht weiter ausgef"uhrt wird und dass sie nicht -- wie manchmal behauptet -- 
auf spezielle Systeme (etwa das Elektron) eingeschr"ankt wird, sondern 
vielmehr universell gelten soll. Allerdings merkt de\,Broglie sogleich 
selbst an, dass diese Hypothese schnell auf  folgenden Widerspruch 
f"uhrt: Nach der Speziellen Relativit"atstheorie m"usste der periodische 
Vorgang eines \emph{bewegten} Objekts auf Grund der Zeitdilatation verlangsamt 
erscheinen. Andererseits sollte die Frequenz des bewegten Objekts aber auch 
h"oher sein, denn seine Energie ist um die kinetische Energie vermehrt. 
De\,Broglie l"oste diesen Widerspruch, indem er zwischen zwei Frequenzen 
unterschied: Der des hypothetischen "`inneren Vorgangs"' und der einer 
den ganzen Raum erf"ullenden Materiewelle, die "`"au"seren"' Bewegung 
des Objekts zugeordnet werden kann. Nur letztere erf"ullt die Bedingung, 
dass ihre Frequenz der Gesamtenergie (mit Einschluss der kinetischen und 
potentiellen) direkt proportional ist. Genau dies wird durch die zwei Jahre 
sp"ater entwickelte Schr"odinger-Gleichung allgemein mathematisch beschrieben. 
Gem"a"s ihr ist, wie bereits oben angemerkt, ein beobachtbarer periodischer 
Vorgang mit einem einzelnen Atom nur dann verbunden, wenn dieses Atom in 
einem Superpositionszustand von Energie-Eigenzust"anden ist. Die dabei 
beobachteten Frequenzen sind dabei stets proportional zu \emph{Differenzen} 
von m"oglichen Energiewerten, die das Atom nach den Gesetzen der 
Quantenmechanik einzunehmen vermag. 

\begin{figure}
\begin{center}
\includegraphics[width=0.36\linewidth]{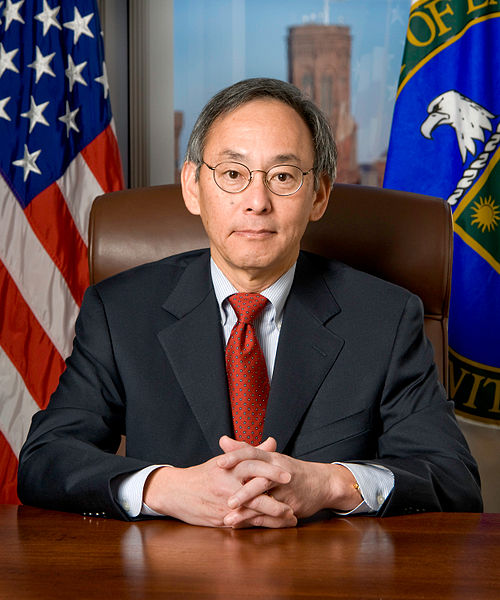}
\hfill
\includegraphics[width=0.45\linewidth]{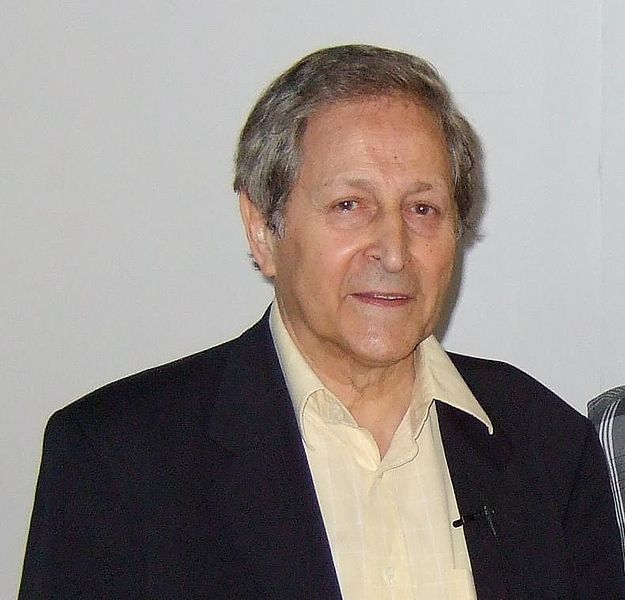}\\
\caption{\small Steven Chu (geb.\,1948) und Claude Cohen-Tannoudji (geb\,1933)}
\end{center}
\end{figure}
Bemerkenswerterweise haben Proponenten und Opponenten der geschilderten 
Interpretation je einen Nobelpreistr"ager des Jahres 1997 auf ihrer 
Seite: Steven Chu von der Stanford University bzw. und Claude Cohen-Tannoudji
von der \'Ecole Normale Sup\'erieure in Paris. Diese teilten sich zusammen mit 
William D. Phillips den Physik-Nobelpreis des Jahres 1997 "`for development 
of methods to cool and trap atoms with laser light"'. Der Proponent, Steven 
Chu, war von Januar 2009 bis April 2013 \emph{Secretary of Energy} der Vereinigten Staaten von Amerika und ist nach wie vor wissenschaftlich aktiv.

\section{Ausblick}
Die Schnittstelle zwischen Gravitationstheorie und Quantenmechanik wirft 
hochinteressante und schwierige konzeptuelle Fragen auf, die beide Theorien 
gleicherma"sen betreffen. Dabei haben wir es bisher nur mit der denkbar 
einfachsten Situation zu tun, n"amlich einem quantenmechanischen System 
in einem \emph{"au"seren} Gravitationsfeld, etwa dem der Erde. Letzteres wird 
klassisch beschrieben, so dass von Quantengravitation im eigentlichen Sinne 
bisher keine Rede ist. 
Der n"achste Schritt best"unde in der Untersuchung der Frage, wie das vom 
System selbst erzeugte Gravitationsfeld beschrieben werden kann und wie es 
auf die Quantendynamik seiner Quellen zur"uckwirkt. Man kann vermuten, dass 
auch diese Frage, zumindest in einem beschr"ankten Rahmen, ohne Kenntnis 
einer vollen Theorie der Quantengravitation zu beantworten ist, so wie 
die elektromagnetische Wechselwirkung quantenmechanischer Systeme zum Teil 
mit klassisch beschriebenen Feldern gelingt. Allerdings treten hier erneut 
die charakteristischen Unterschiede der gravitativen zu anderen Wechselwirkung 
auf und fordern wieder dazu auf, grunds"atzliche Fragen frisch zu 
"uberdenken. Eine der immer wieder diskutierten m"oglichen Konsequenzen 
gravitativer Selbstwechselwirkung ist eine nicht-lineare Modifikation 
der Schr"odinger-Gleichung. Diese f"uhrt zu einer Verlangsamung der 
Dispersion und damit zu einer m"oglichen Abschw"achung von Interferenzeffekten 
von Materiewellen, die gem"a"s numerischer und analytischer Absch"atzungen 
zwar nicht in gegenw"artigen, aber m"oglicherweise in zuk"unftigen 
Experimenten dargestellt werden k"onnten. Ein solcher Nachweis b"ote nicht 
nur die ersten direkten Einblicke in die gravitative Selbstwechselwirkung 
von Quantensystemen, er k"onnte auch wertvolle Hinweise "uber den Einfluss
geben, den die Gravitation auf unser physikalisches Verst"andnis der Quantenphysik 
nimmt. Alle letztlich vergeblich gebliebenen bisherigen Versuche, eine 
konsistente Theorie der Quantengravitation zu formulieren, gehen davon aus, 
dass "`Quantisierungsregeln"', die sich im Zusammenhang mit anderen 
Wechselwirkungen bew"ahrt haben, auch mehr oder weniger angepasst auf die 
Gravitation "ubertragen lassen. Ein physikalisches Verst"andnis der dabei 
implizit gemachten Annahmen ist aber so gut wie nicht vorhanden. An dieser 
Stelle kann das hier geschilderte Vorgehen konstruktive Beitr"age liefen, nicht 
zuletzt wegen der realistischen Aussicht auf experimentelle Entscheidbarkeit 
der aufgeworfenen Fragen durch die anhaltend rasante Entwicklung der 
Interferometrie mit Materiewellen. Auch damit w"are man zwar immer noch weit 
vom eigentlichen Kerngebiet der Quantengravitation entfernt, darf aber dennoch 
hoffen, wertvolle Hinweise auf die notwendigen begrifflichen Anpassungen zu 
bekommen, nach denen die Forschung auf dem Gebiet der Quantengravitation so 
h"anderingend sucht und denen innerhalb der Grenzen der Phantasie mit dem 
alleinigen Kriterium der formalen Konsistenz augenscheinlich nicht beizukommen 
ist. 

Derzeit wird intensiv an der Optimierung der experimentellen Rahmenbedingungen 
zur Steigerung der Genauigkeiten gearbeitet, z.B. an der Verl"angerung der 
Freiflugzeiten im Interferometer. Wie wir oben sahen, ist die 
Phasenverschiebung  im Atom\-interferometer proportional zum Quadrat dieser 
Zeit aber nur direkt proportional zur Gravitationsbeschleunigung. Da die 
Freiflugzeit auch durch die Gravitationsbeschleunigung begrenzt wird, kann 
es also g"unstig sein, letztere  durch einen kontrollierten Fall in einem 
Fallturm zu verringern, oder das Experiment in einem Satelliten 
unterzubringen, um so den quadratischen Effekt der erh"ohten Freiflugzeit 
auszunutzen. Man darf also auf die Entwicklung der n"achsten 10 Jahre 
sehr gespannt sein, die es erlauben wird, fundamentale Fragestellungen 
zum Verh"altnis von Gravitation und Quantentheorie innerhalb machbarer 
Physik zu adressieren.    

\subsection*{Danksagung}
Ich danke dem Exzellenzcluster QUEST (Centre for Quantum Engineering and 
Space-Time Research, www.questhannover.de) und den beteiligten Kollegen 
an der Leibniz Universit"at Hannover sowie dem ZARM in Bremen f"ur 
Unterst"utzung und viele wertvolle Gespr"ache im Umkreis dieses 
spannenden Themas. 

%% Hier kommen die Kaesten
%%%%%%%%%%%%%%%%%%%%%%%%%%%

\newpage
\section*{Anh"ange}

\subsection*{1. Die Planck Einheiten}
%\label{Kasten:PlanckEinheiten}
%
Aus der Lichtgeschwindigkeit 
($c=299\,792\,458\,\mathrm{m\cdot s^{-1}}$, 
der Gravitationskonstante 
$G=6{,}673\times 10^{-11}\,\mathrm{m^3\cdot kg^{-1}\cdot s^{-2}}$
und dem Planck'schen Wirkungsquantum 
($h=6{,}626\times 10^{-34}\,\mathrm{kg\cdot m^2\cdot s^{-1}}$)
lassen sich die Planck'sche L"ange $L_P$, die Planck'sche
Zeit $T_P$ und die Planck'sche Masse $M_P$ ableiten
\begin{subequations}
\label{eq:PlanckSkalen}
\begin{alignat}{4}
\label{eq:PlanckLaenge}
&L_P&&\,=\,\sqrt{\frac{h\cdot G}{c^3}}&&\,=\,4{,}05\times 10^{-35}\ &&\mathrm{Meter}\,,\\
\label{eq:PlanckZeit}
&T_P&&\,=\,\sqrt{\frac{h\cdot G}{c^5}}&&\,=\,1{,}35\times 10^{-43}\ &&\mathrm{Sekunden}\,,\\
\label{eq:PlanckMasse}
&M_P&&\,=\,\sqrt{\frac{h\cdot c}{G}}&&\,=\,5{,}46\times 10^{-5}\ &&\mathrm{Gramm}\,.
\end{alignat}
\end{subequations}
Die L"ange $L_p$ gibt die Antwort auf die etwas naive Frage, 
unterhalb welcher Wellenl"ange ein einzelnes Licht- oder 
Materie-Quant genug Eigengravitation bes"a"se um ein 
schwarzen Loch zu bilden. Dieses h"atte dann eine Masse von 
$M_p$ oder dar"uber. Schwarze L"ocher mit Massen unterhalb 
$M_P$ w"aren sicher nicht mehr klassisch beschreibbar, denn 
ihre Comptonwellenl"ange w"are gr"o"ser als ihr Horizontradius. 

%\newpage
\subsection*{2. Langreichweitige Wechselwirkungen 
und UFF-Verletzungen}
%\label{Kasten:LangreichweitigeWW}
%
Das Gravitationspotential eines Testk"orpers $A$ der Masse $M_A$ 
im Gravitationsfeld der Erde, $E$, deren Masse mit $M_E$ bezeichnet 
sei, ist 
\begin{equation}
\label{eq:Gravitationspotential}
V_{AE}=-\,G\,\frac{M_AM_E}{r_{AE}}\,.
\end{equation}
Dabei ist $r_{AB}$ der Abstand zwischen Erde und Testk"orper
(d.h. ihrer Schwerpunkte). Eine weitere langreichweitige Kraft, 
die an die Ladungen $Q_A$ und $Q_E$ der K"orper $A$ und $E$ 
koppelt, w"urde einem weiteren Potential der Form   
\begin{equation}
\label{eq:Q-Potential}
W_{AE}=-\,H\,\frac{Q_AQ_E}{r_{AE}}
\end{equation}
entsprechen, wobei $H$ die Kopplungskonstante der entsprechenden 
Wechselwirkung ist. Wir beschr"anken uns hier auf den relevanten 
Fall eines langreichweitigen Skalarfeldes, in dem die Kraft also 
anziehend und das zugeh"orige Potential negativ ist. (F"ur ein 
Vektorfeld w"are die Kraft absto"send und das Potential entsprechend 
positiv anzusetzen. W"are die Kraft nicht langreichweitig und das
Austauschteilchen entsprechend mit einer nicht-verschwindenden 
Ruhemasse versehen, so w"urde das Potential und damit die Kraft 
mit dem Abstand zus"atzlich exponentiell unterdr"uckt werden.) 

Das Gesamtpotential ist die Summe der beiden. Diese Summe hat 
mathematisch wieder die Form eines einzigen Gravitationspotentials,
allerdings mit einer von den K"orpern abh"angigen (also nicht mehr universeller) Gravitationskonstante:
\begin{equation}
\label{eq:TotalesPotential-1}
U=V+W=-\,G_{AE}\,\frac{M_AM_B}{r_{AB}}\,,
\end{equation}
wobei  
\begin{equation}
\label{eq:TotalesPotential-2}
G_{AE}=G\bigl(1+h\,q_Aq_E\bigr)\,.
\end{equation}
Dabei ist zur Abk"urzung $h=H/G$ gesetzt und $q_A=Q_A/M_A$ bezeichnet 
die auf die die Masseneinheit bezogene Ladung (spezifische Ladung). 
Alternativ kann man diesen Sachverhalt auch durch die Ungleichheit von
tr"ager und schwerer Masse ausdr"ucken. Ist n"amlich die spezifische 
Ladung f"ur zwei verschiedene K"orper $A$ und $B$ unterschiedlich, 
so erfahren diese in Feld der Erde auch unterschiedliche Beschleunigungen. 
Sind die Abweichungen von $G_{AE}$ und $G_{BE}$ zu $G$ klein, dann 
sieht man leicht, dass der E"otv"os-Faktor in f"uhrender Ordnung gegeben 
ist durch  
\begin{equation}
\label{eq:EoetvoesFactor}
\eta(A,B)\approx hq_E(q_A-q_B)\,,
\end{equation}
also proportional zur Differenz der spezifischen Ladungen.

\newpage
\bibliographystyle{plain}
\bibliography{RELATIVITY,HIST-PHIL-SCI,MATH,QM,COSMOLOGY}

\begin{thebibliography}{10}

\bibitem{Chou.EtAl:2010}
Chin-Wen Chou, David Hume, Till Rosenband, and David Wineland.
\newblock Optical clocks and relativity.
\newblock {\em Science}, 329(5999):1630--1633, 2010.

\bibitem{Damour.Polyakov:1994a}
Thibault Damour and Polyakov Alexander.
\newblock String theory and gravity.
\newblock {\em General Relativity and Gravitation}, 26(12):1171--1176, 1994.

\bibitem{Fray.Weitz:2009}
Sebastian Fray and Martin Weitz.
\newblock Atom-based tests of the equivalence principle.
\newblock {\em Space Science Reviews}, 148:225--232, 2009.

\bibitem{Giulini:2012}
Domenico Giulini.
\newblock Equivalence principle, quantum mechanics, and atom-interferometric
  tests.
\newblock In F.~Finster, O.~M\"uller, M.~Nardmann, J.~Tolksdorf, and
  E.~Zeidler, editors, {\em Quantum {F}ield {T}heory and {G}ravity. Conceptual
  and mathematical advances in the search for a unified framework}.
  Birkh\"auser Verlag, Basel, 2012.
\newblock arXiv:1105.0749v2.

\bibitem{Greenberger.Overhauser:1980}
Daniel~M. Greenberger and Albert~W. Overhauser.
\newblock {Gravitation und Quantentheorie}.
\newblock {\em Spektrum der Wissenschaft}, pages 42--57, Juli 1980.

\bibitem{Herrmann.etal:2009}
Sven Herrmann et~al.
\newblock Rotating optical cavity experiment testing {Lorentz} invariance at
  the $10^{-17}$ level.
\newblock {\em Physical Review~D}, 80(10):105011 (8 pages), 2009.

\bibitem{Hertz:ConstitutionDerMaterie}
Heinrich Hertz.
\newblock {\em {Die Constitution der Materie: Eine Vorlesung \"uber die
  Grundlagen der Physik aus dem Jahre 1884}}.
\newblock Springer Verlag, Berlin, 1999.

\bibitem{Mueller.etal:2010}
Holger M\"uller, Achim Peters, and Steven Chu.
\newblock A precision measurement of the gravitational redshift by the
  interference of matter waves.
\newblock {\em Nature}, 463:926--930, 2010.

\bibitem{Nesvizhevsky:Nature-2002}
Valery~V. Nesvizhevsky et~al.
\newblock Quantum states of neutrons in the {Earth's} gravitational field.
\newblock {\em Nature}, 415(17 January):297--299, 2002.

\bibitem{Peters.Chung.Chu:1999}
Achim Peters, Keng~Yeow Chung, and Steven Chu.
\newblock Measurement of gravitational acceleration by dropping atoms.
\newblock {\em Nature}, 400(6747):849--852, 1999.

\bibitem{vanZoest:2010}
Tim van Zoest et~al.
\newblock {Bose-Einstein} condensation in microgravity.
\newblock {\em Science}, 328(5985):1540--1543, 2010.

\end{thebibliography}
\end{document}